\documentclass[twocolumn,trackchanges]{aastex6}
\usepackage{color,amsmath,textcomp,url,graphicx,subfigure,latexsym}
\usepackage{epsfig}
\usepackage{amssymb}%
\usepackage{longtable}%
\usepackage{rotating}

\shorttitle{}
\shortauthors{Kellogg et al.}

\begin{document}

\title{Characterizing The Cloud Decks of Luhman 16AB with Medium-Resolution Spectroscopic Monitoring}
\author{Kendra Kellogg\altaffilmark{1}, Stanimir Metchev\altaffilmark{1,2}, Aren Heinze\altaffilmark{3}, Jonathan Gagn\'e\altaffilmark{4,5}, Radostin Kurtev\altaffilmark{6,7}}
\altaffiltext{1}{The University of Western Ontario, Centre for Planetary and Space Exploration, 1151 Richmond St, London, ON N6A 3K7, Canada; kkellogg@uwo.ca, smetchev@uwo.ca}
\altaffiltext{2}{Stony Brook University, Department of Physics \& Astronomy, 100 Nicolls Rd, Stony Brook, NY 11794-3800, USA}
\altaffiltext{3}{Institute for Astronomy, 2680 Woodlawn Drive, Honolulu, HI 96822-1839, USA}
\altaffiltext{4}{Carnegie Institution of Washington DTM, 5241 Broad Branch Road NW, Washington, DC~20015, USA}
\altaffiltext{5}{NASA Sagan Fellow}
\altaffiltext{6}{Instituto de F\'{i}sica y Astronom\'{i}a, Facultad de Ciencias, Universidad de Valpara\'{i}so, Ave$.$ Gran Breta\~{n}a 1111, Playa Ancha, Casilla 53, Valpara\'{i}so, Chile}
\altaffiltext{7}{Millennium Institute of Astrophysics, Chile}

\begin{abstract}
We present results from a two-night $R\sim4000$ 0.9--2.5\micron\ spectroscopic monitoring campaign of Luhman 16AB (L7.5 + T0.5). We assess the variability amplitude as a function of pressure level in the atmosphere of Luhman 16B: the more variable of the two components. The amplitude decreases monotonically with decreasing pressure, indicating that the source of variability---most likely patchy clouds---lies in the lower atmosphere. An unexpected result is that the strength of the \ion{K}{1} absorption is higher in the faint state of Luhman 16B and lower in the bright state. We conclude that either the abundance of \ion{K}{1} increases when the clouds roll in, potentially because of additional \ion{K}{1} in the cloud itself, or that the temperature-pressure profile changes. We reproduce the change in \ion{K}{1} absorption strengths with combinations of spectral templates to represent the bright and the faint variability states.  These are dominated by a warmer L8 or L9 component, with a smaller contribution from a cooler T1 or T2 component.  The success of this approach argues that the mechanism responsible for brown dwarf variability is also behind the diverse spectral morphology across the L-to-T transition.  We further suggest that the L9--T1 part of the sequence represents a narrow but random ordering of effective temperatures and cloud fractions, obscured by the monotonic progression in methane absorption strength.
\end{abstract}

\keywords{brown dwarfs --- stars: individual (WISE J104915.57--531906.1AB)}

\footnote{This paper includes data gathered with the 6.5 meter Magellan Telescopes located at Las Campanas Observatory, Chile.}

\section{Introduction}

The spectral morphology of brown dwarfs changes drastically ($\sim$6 subtypes) over a relatively narrow range of effective temperatures ($\sim$1200--1300K) at the transition from spectral type L to T. During this transition, clouds in the atmospheres of mid-L objects disappear and by mid-T spectral types the visible atmospheres are cloud-free. It is not yet well understood how these clouds disappear or if it is possible that objects skip spectral types rather than evolve through all intervening spectral morphologies.

A characteristic of ultracool atmospheres in this spectral type range is flux variability on the rotation period of the object \citep{radigan13,radiganetal14}.  Summarizing the evidence from ground-based photometric monitoring campaigns in the $J$ band, \cite{radigan14} finds that large-amplitude ($>$2\%) variability is common for brown dwarfs spanning the L-to-T transition.  Among L9--T3.5 dwarfs $24\%^{+11\%}_{-9\%}$ are large-amplitude variables, vs.\ only $3\%^{+3\%}_{-2\%}$ among L0--L8.5 and T4--T9.5 dwarfs.  The near-infrared (NIR) 1--2.4~$\micron$ variability has been attributed to clear holes in a prevailing cloud deck that reveal deeper, hotter parts of the atmosphere (e.g., \citealp{artigau09}) or, more generally, to multiple patchy cloud layers that lead to net changes in the observed photospheric level as a brown dwarf rotates \citep{apai13}.   The picture of dissipating cloud decks across the L-to-T transition is consistent with the rainout of condensates invoked in the progression from dusty L-type to clear T-type atmospheres \citep{ackerman01,burgasser02}.

Longer-wavelength 3--5~$\micron$ monitoring during the \textit{Weather on Other Worlds} program with the \textit{Spitzer Space Telescope} has now revealed that smaller-amplitude variations are likely ubiquitous in L and T dwarfs across all spectral types \citep{metchev15}.  That study finds that $80\%^{+20\%}_{-27\%}$ of L3--L9.5 dwarfs vary by $>$0.2\% and $36\%^{+26\%}_{-17\%}$ of T0--T8 dwarfs vary by $>$0.4\% in the \textit{Spitzer} [3.6] and [4.5] bands.  Unlike \cite{radigan14}, \cite{metchev15} find no evidence for enhancement in either frequency or amplitude of variability at the L/T transition, but instead observe that the maximum detected amplitudes increase throughout the L3--T8 range.  The different dependence of variability on spectral type measured at $JHK$ (1--2.4 \micron) relative to the \textit{Spitzer} IRAC bands (3--5 \micron) is likely rooted in the difference among the atmospheric pressure levels---and correspondingly, sources of opacity---observed at these wavelengths.   NIR observations, especially in the $J$ band, can be sensitive to deep condensate clouds at $\sim$10~bar pressures, while 3--5~$\micron$ photometry probes mostly upper-atmosphere carbon monoxide and methane gas opacity at 0.1--1~bar (e.g., \citealp{ackerman01}). 

To test this hypothesis, we need simultaneous observations with a broad wavelength range that probe multiple pressure levels. Precise low-dispersion ($R\sim130$) 1.1--1.6~$\micron$ spectrophotometric monitoring with WFC3 on the \textit{Hubble Space Telescope} of Luhman 16AB \citep{luhman13} shows that there are multiple photospheric regions involved in the flux variations \citep{apai13,buenzli15a}, possibly with different brightness temperatures \citep{karalidi16}.  \cite{apai13} demonstrate that the photospheric inhomogeneities can be attributed to regions with various contributions from cool thick clouds and warmer, brighter, and thinner clouds. A similar conclusion is reached by \cite{burgasser14} from a $R\sim120$ resolved spectroscopic sequence on a binary from ground-based observations with SpeX on the IRTF.  Multiple spots can also produce phase lags among the light curves at different wavelengths, as have already been reported from simultaneous 1--5~$\micron$ \textit{Hubble} and \textit{Spitzer} observations of L and T dwarfs \citep{apai13,yang16}.

Higher-dispersion ($R>1000$) monitoring can be used to accurately estimate the altitudes of the various cloud decks.  Given sufficient spectral resolution, the line profiles of gravity-sensitive absorbers, such as alkali elements, can be resolved throughout the optical and the NIR.  Variations in the strengths of these lines, corresponding to holes in the upper cloud deck rotating in and out of view, can thus reveal the pressure levels where cloud condensates form.   

The nearest pair of brown dwarfs---the L7.5 + T0.5 binary WISE J104915.57--531906.1AB (a.k.a., Luhman 16AB; \citealp{luhman13})---offers one of the best opportunities for high dispersion spectroscopy of substellar objects.  Both components are bright (MKO $J$ magnitudes of 11.5 and 11.2 for the A and B components, respectively; \citealp{burgasser13}), and spatially well resolved ($1\farcs5$ at the time of discovery; \citealp{luhman13}).  Photometric monitoring has shown that the secondary T0.5 component can vary by upwards of 10\% peak-to-peak in the optical through NIR \citep{gillon13,biller13,burgasser14,buenzli15a}.  \cite{biller13} found that the variations of Luhman 16B in the bandpasses that probe the lower atmosphere ($z^\prime$, $J$, and $H$) are in phase while the variations in the bands that probe the upper atmosphere ($r^\prime$, $i^\prime$ and $K$) are out of phase by $\geq$100$^\circ$ with respect to the $z^\prime$ band.   Since then, WFC3/\textit{HST} observations by \cite{buenzli15b} have shown that the primary L7.5 component also varies, with a peak-to-peak amplitude of up to 4.5\%.  The variability periods of the components are 5.05~h for component B, and either 5~h or 8~h for component A \citep{karalidi16}.  The periodic variations in the binary components can thus be conveniently studied over the course of a single observing night. 

Luhman~16AB has already been intensively investigated at high spectral resolution.  Most notably, \cite{crossfield14} used very high dispersion ($R=50,000$) spectroscopy to invert the time-dependence in the Doppler broadening of the molecular line profiles to produce a spatially resolved map of brightness variations of Luhman 16B.   A comparative medium resolution ($R\sim4000$) 0.8--2.4~$\mu$m analysis of the binary by \cite{faherty14} found that the T0.5-type B component shows stronger absorption in the gravity-sensitive 1.168, 1.177, 1.243, and 1.254~$\mu$m \ion{K}{1} lines than the L7.5-type A component, while having a brighter continuum in the 1.17~$\mu$m and 1.25~$\mu$m region.  They interpret this as a verification of the cloud dissipation hypothesis across the L/T transition: the cooler component has a patchy upper cloud layer that reveals the hotter, higher-pressure inner parts of the atmosphere.  Where the molecular opacity is low, such as in the $J$ band, these make the B component brighter than the hotter L7.5-type A component of the binary.  The \cite{faherty14} analysis does not address the variability of the Luhman~16 components, as it combines altogether only 40 min of exposure on each component.  

Here we present a two night-long $R\sim4000$, 0.9--2.5 $\micron$ spectroscopic monitoring study of Luhman 16AB.  We analyze the behavior of the \ion{K}{1}, CH$_4$, FeH, and H$_2$O absorption strengths in order to determine the nature of the variability phenomenon.  A companion study (Heinze et al. 2017, in preparation) will describe a complementary and contemporaneous set of $R\sim300$ optical spectra, and will investigate the variability of the much stronger \ion{K}{1} 0.77~$\micron$ absorption doublet.

\section{Observations and Data Reduction}

We obtained simultaneous observations of Luhman 16AB with the Folded-port InfraRed Echellette spectrograph (FIRE; \citealp{simcoe08,simcoe13}) on the Magellan, Baade telescope and the Gemini Multi-Object Spectrometer (GMOS) on the Gemini-South telescope on 2014, February 23 and 24. Both nights were clear with a stable seeing between 0.8 and 1 arcsec. We used the 1$\farcs$0 slit on FIRE in cross-dispersed mode ($R\sim4000$) covering 0.9--2.5\micron.  Observations were taken over two nights for a monitoring of $\sim$2 rotations of Luhman 16B per night. We only report the results from the NIR observations here. The results from the optical observations will be discussed in Heinze et al. (2017). Because we monitored Luhman 16AB over complete nights, our observations likely inherit cyclic correlations with seeing and airmass.  We discuss methods to mitigate these effects in the calibration of our spectra in $\S$\ref{sec:calib}.

A total of forty-six 600s exposures were taken with an ABBA nodding sequence with the slit aligned on both components for simultaneous observations of Luhman 16 A and B. Several observations of a telluric standard, HD~98042, were taken throughout the observing sequence on both nights. We used gain settings of 3.8 e$^-$/DN for the science and the standard observations. Illumination and appropriate pixel flats were observed either at the beginning or the end of the night and a neon-argon lamp was observed immediately after each set of target and standard star observations for use in instrumental calibrations. All science and telluric observations were taken using the sample-up-the-ramp (SUTR) readout mode.

We reduced the data using the Interactive Data Language (IDL) pipeline FIREHOSE, which is based on the MASE \citep{bochanski09} and SpeXTool \citep{vacca03, cushing04} packages. A majority of the bad pixels were removed initially using a bad pixel mask. The pipeline was modified to subsequently achieve a better rejection of bad pixels and includes a barycentric velocity correction and converts the wavelength solution to vacuum. The extraction profiles built by FIREHOSE cannot account for traces of both objects at once, hence it is only an approximate profile that covers both traces in the case of science exposures. The profiles were used only for the purpose of straightening the 2D traces in a first step using a custom Interactive Data Language (IDL) routine. In a second step, the two-dimensional wavelength solutions were used to un-shear the two-dimensional raw exposures in the wavelength direction. These steps resulted in a straight horizontal spectral trace with a uniform wavelength solution on every row for every exposure and order. 

A custom IDL pipeline was then used to extract the spectral traces of the two Luhman 16 AB components separately. At each column of the two-dimensional exposures, two Moffat functions \citep{moffat69} were fit simultaneously to the data using the {\sc mpfitfun.pro} routine, which performs a Levenberg-Marquardt least-squares fit. The resulting area under the curve of the individual best-fitting Moffat profiles then provide the separate raw spectral flux densities of Luhman 16 A and B. We calculate that an average of $\sim$10\% of the measurement error is attributed to systematic flux contamination from the opposite component. A similar one-profile algorithm was used to extract the telluric standard, which was then used to correct telluric absorption from the Earth's atmosphere using the {\sc fire\_xtellcor.pro} IDL routine.

\begin{figure*}
\centering 
\includegraphics[scale=0.4]{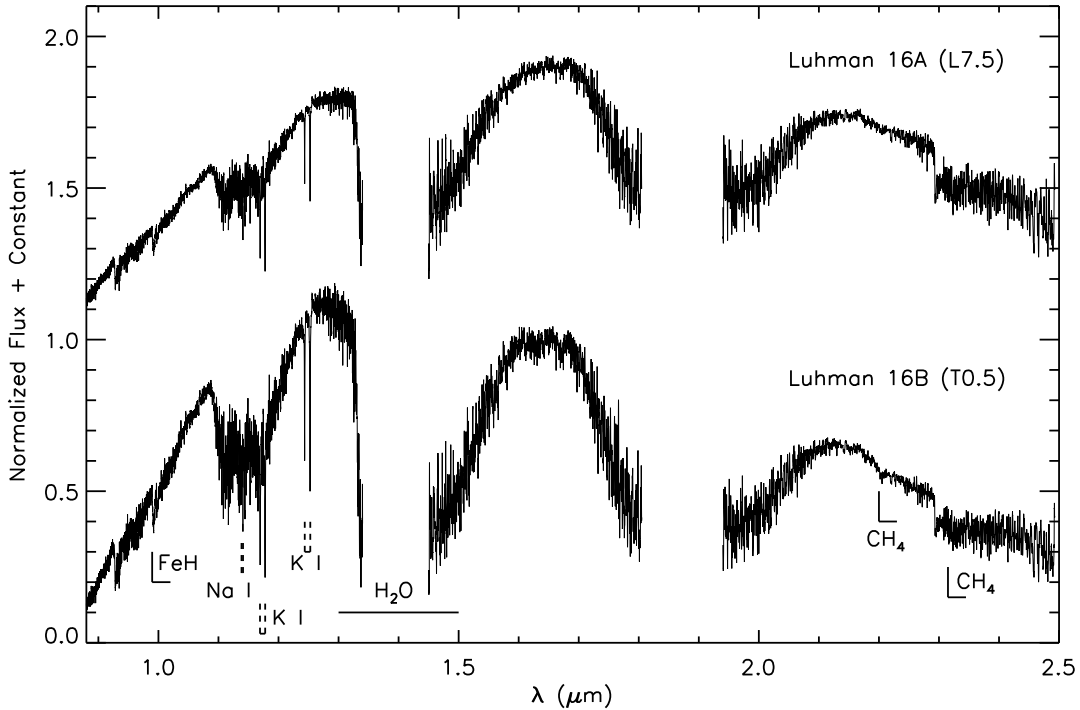}
\caption{\footnotesize The $R\sim4000$ Magellan FIRE spectra of the Luhman 16A and B components from 2014, February 23 and 24. The spectra for each object were median-combined order by order and the orders were subsequently stitched.}
\label{fig:1dcomb}
\end{figure*}

Figure \ref{fig:1dcomb} shows spectra of the two components---Luhman 16A (L7.5) and Luhman 16B (T0.5)---median-combined from both nights of observations (6 hr total exposure). 

\section{Analysis and Results}\label{sec:res}

Our goal was to understand how the strengths of spectroscopic features, particularly pressure-sensitive lines, change as Luhman 16B varies. We first discuss how we calibrated the light curves of Luhman 16B with Luhman 16A to remove systematic effects. Then we discuss how we created spectra that represented the bright and faint states of Luhman 16B to assess changes in the strengths of the absorption features between states. Finally, we discuss the variations and changes we observed in the spectra.

\subsection{Light Curve Calibration}\label{sec:calib}

\begin{figure*}
\centering 
\includegraphics[scale=0.3]{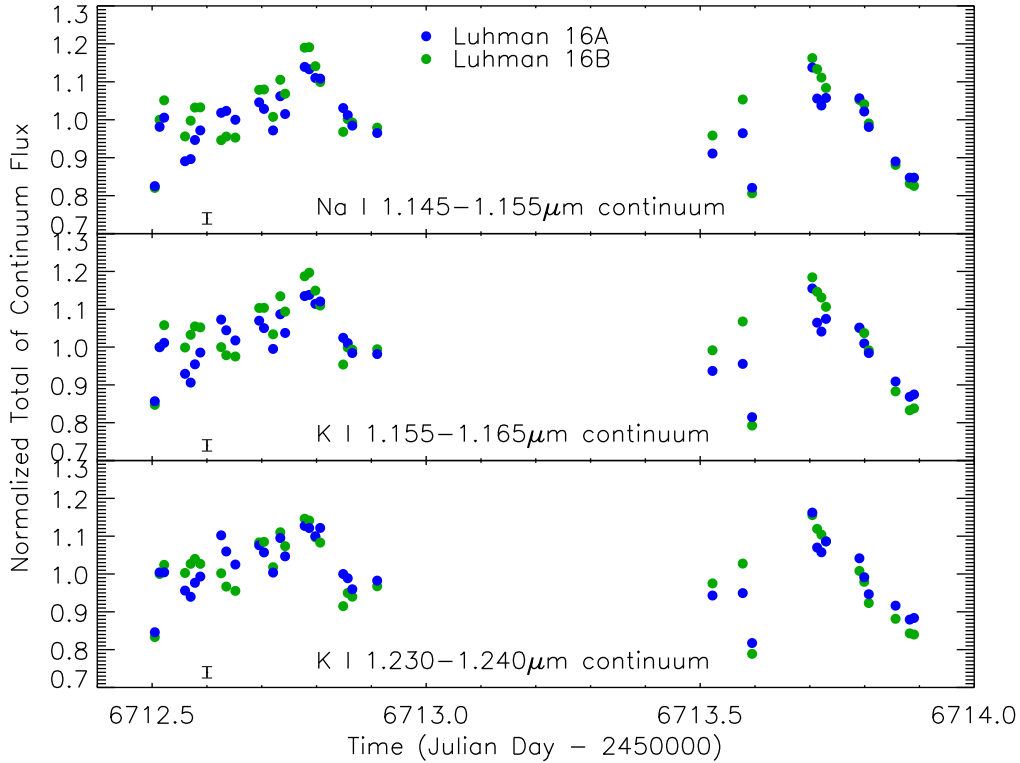}
\caption{\footnotesize Continuum variations of Luhman 16A and B (blue and green circles, respectively) around the \ion{Na}{1} and \ion{K}{1} NIR lines from the two nights of FIRE observations. The data have been normalized to the respective mean values of the entire observing period. Wavelength regions for the summation of the continuum fluxes are 1.145--1.155\micron\ for the \ion{Na}{1} doublet and 1.155--1.165\micron,  1.230--1.240\micron\ for the \ion{K}{1} doublets; i.e., excluding the doublets themselves. The average random uncertainties are shown in the bottom left corner.}
\label{fig:lightcurve}
\end{figure*}

Figure \ref{fig:lightcurve} shows the sum of the continuum flux in the orders of the pressure-sensitive alkali doublets (1.145--1.155\micron, 1.155--1.165\micron, and 1.230--1.240\micron) for each spectrum taken over the span of two nights. It is immediately evident that the fluxes in the respective orders of A and B correlate strongly, which we attribute to seeing and slit-loss variations during the monitoring sequence.  Because components A and B were aligned along the slit, these variations are correlated.  We eliminated these relative systematics and obtained a corrected light curve of the variability of the T0.5 dwarf B component by dividing the spectra of component B by those of component A.

Although the L7.5 dwarf A component is also known to vary with a $\gtrsim$2 times smaller amplitude \citep{buenzli15b}, our primary concern was the elimination of the much larger-amplitude variations caused by the seeing.  Relative calibration with respect to A was the only option to achieve this with a long-slit spectrograph.  The GMOS optical spectra of component B were also normalized by the A component (Heinze et al.\ 2017), therefore, using a ratio of the FIRE spectra of the components allows a direct comparison between the variability in the NIR and in the optical. 

\begin{figure*}
\centering 
\includegraphics[scale=0.275]{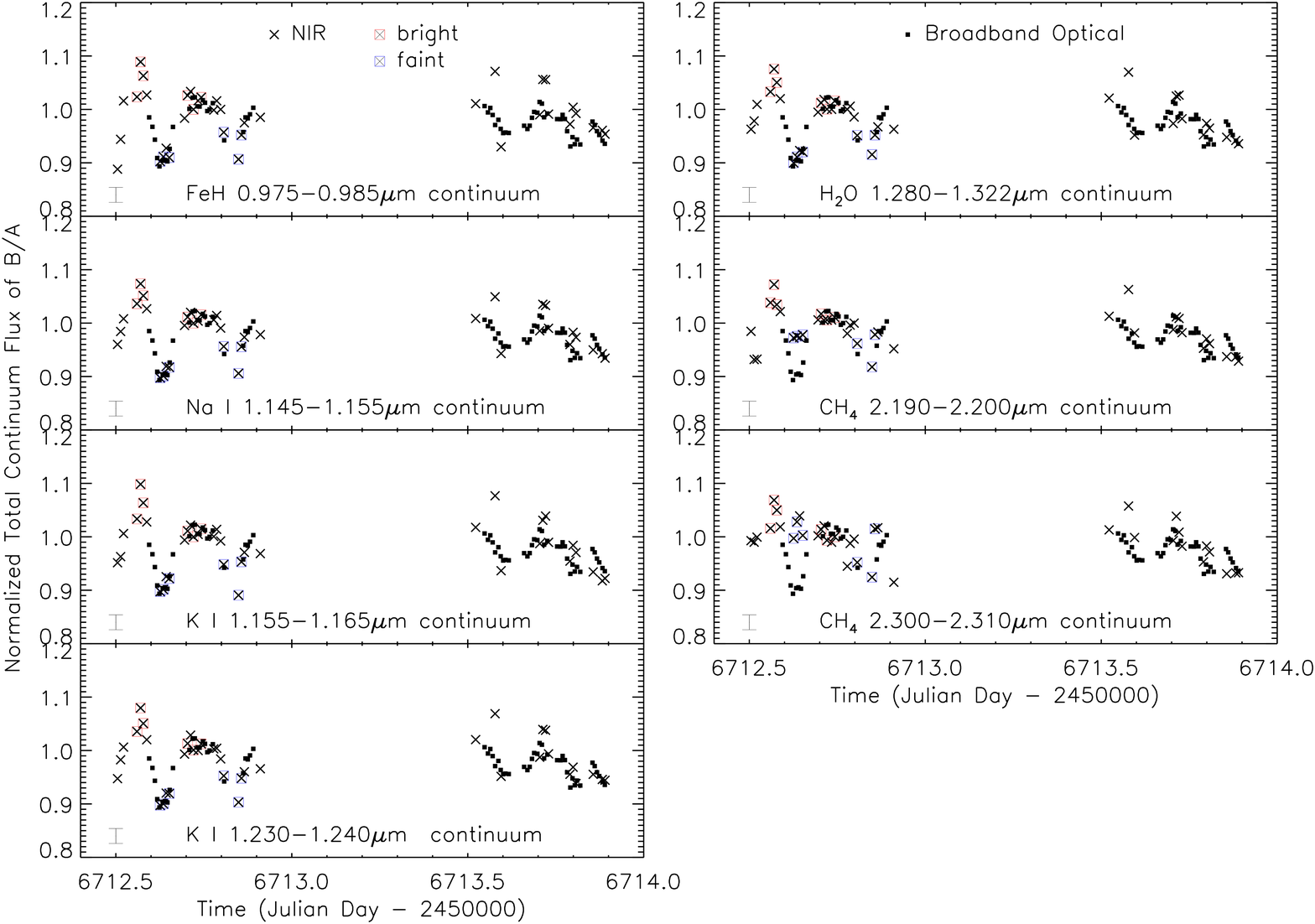}
\caption{\footnotesize Variations in the continuum around the various absorption features of the ratio of Luhman 16B/A from the two nights of FIRE observations. The GMOS 0.8--1.0$\micron$ light curve is also shown for comparison. The data are normalized to the respective mean values of the observations that overlapped in time. The crosses are the NIR data and the squares are the optical data (Heinze et al.\ 2017). The points corresponding to each of the six spectra used for the bright and faint states are indicated in red and blue, respectively. The average total uncertainties are shown in the bottom left corners.}
\label{fig:lightcurve_ratio}
\end{figure*}

We show the thus-normalized continuum flux variations in three of the panels of Figure \ref{fig:lightcurve_ratio}.  We have also applied the same technique to other wavelength regions of interest: namely the continua around FeH (0.975--0.985\micron), H$_2$O (1.280-1.320\micron) and CH$_4$ (2.190--2.200\micron\ and 2.300--2.310\micron), and show these in the other panels of Figure \ref{fig:lightcurve_ratio}. We see from this Figure that the exposures of the optical and NIR data are not entirely synchronized because of the logistics of the individual observations.  In particular, time was lost at the beginning of the first night of GMOS observations because of a technical fault. The most prominent peak and trough of the light curve and the clearest trend of variability occur at the beginning of the two-night observing period for both the NIR and optical data.  

\subsection{Creation of Representative Bright and Faint State Spectra}\label{sec:ratio}

We are particularly interested in how the relative strengths of the absorption features change over the course of the observations as this offers insight into the pressure level where the variability arises. To assess this, we identified the observations that were taken at the maxima and minima of the light curve (Figure \ref{fig:lightcurve_ratio}). We used the continuum around the $J$-band \ion{K}{1} doublets to identify the minima and maxima, and selected the same sets of spectra for all other wavelength regions for consistency. While we used the light curve of the B/A ratios to identify the bright- and faint-state spectra, we only used the telluric-corrected spectra of the B component when median-combining to create representative bright and faint spectra. We thus avoided contaminating our results with any effects from the variation in the primary A component.

The variations during the second night appeared less organized than during the first night.  Consequently, we were unable to select moments when the conditions contributing to either the bright or to the faint state were clearly expressed. Such changes in the light curve are not unexpected.  Long-term monitoring of SIMP~J013656.5+093347 shows that the light curve becomes irregular at some epochs (e.g., \citealp{artigau09,metchev13}). For this reason, we excluded night two from our analysis.  

To assess the amplitudes of the absorption feature variations in the T0.5 dwarf B component, we formed the ratio of the spectra representing the bright and faint states. The representative maximum and minimum spectra were created by using the three spectra of Luhman 16B that cluster around each of the two peaks of the light curve (bright state) and the three spectra that clustered around each of the two troughs (faint state): for a total of six spectra for each state. We did not normalize these spectra by the corresponding spectra of Luhman 16A: to avoid incurring an unknown correlation with the spectroscopic variability of the primary component.  However, this meant that our selected maximum- and minimum-light spectra of Luhman 16B were again subject to slit losses and changing atmospheric conditions, and so had systematic slope discrepancies.  To mitigate this, we removed any slopes in the continua by filtering out the lowest frequency from a Fast Fourier Transform.  We thus remove sensitivity to overall color changes but retain sensitivity to changes in the higher order features such as absorption lines.  We then scaled the corrected spectra to the average continuum level of the six in each of the maximum- and minimum-light sets.  

\begin{figure*}
\centering 
\includegraphics[scale=0.275]{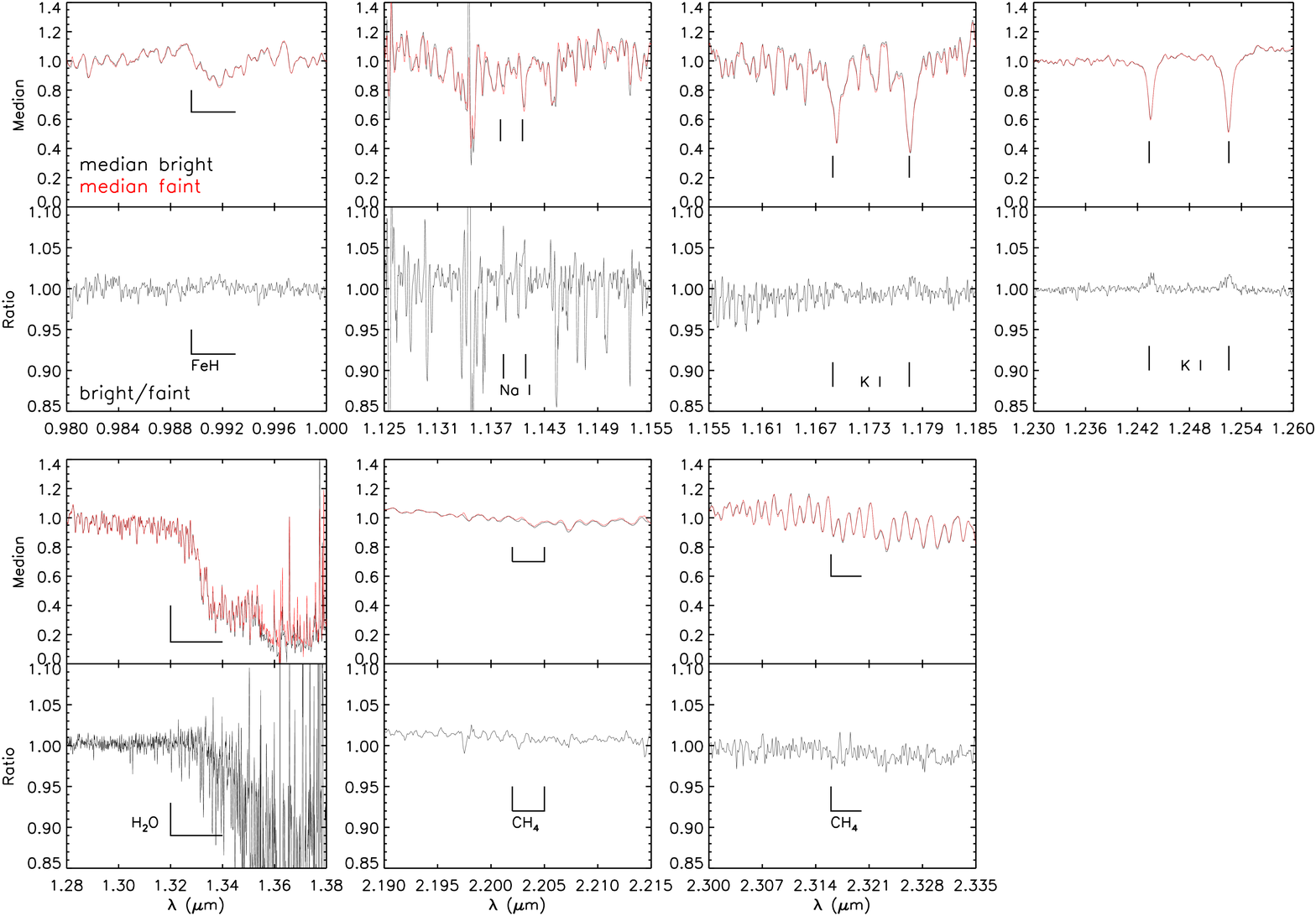}
\caption{\footnotesize Relative absorption strength changes in the spectral regions of interest: around the FeH, \ion{Na}{1}, \ion{K}{1}, H$_2$O, and CH$_4$ absorbers.  The top sets of panels show the representative spectra of Luhman 16B at maximum (black) and minimum (red) light.  The bottom sets of panels show the maximum-to-minimum light spectral ratios. We see changes in the absorption strengths of the two \ion{K}{1} doublets and the water band, but not of the other absorption features.}
\label{fig:fracdiff_n1}
\end{figure*}

We also noticed a slight shift in the wavelength solution for the spectra over the course of the observations which resulted from a lack of frequent observations of an arc lamp. Before combining the spectra, we cross-correlated all the spectra with respect to the first and shifted them by the appropriate amount. Finally, we median-combined and re-normalized the six scaled spectra for each set to unity in the continua, which resulted in two representative spectra: one for the bright and one for the faint state.   While the two representative spectra lack absolute flux calibration, the relative flux information within each of the two spectra is precisely calibrated. The top panels in Figure \ref{fig:fracdiff_n1} show the median-combined representative spectra for the bright (black) and faint (red) states for night one in all of the spectral regions of interest.  We formed ratios of the bright and faint states to assess the relative changes in line strengths: shown in the bottom sets of panels in Figure \ref{fig:fracdiff_n1}. 

\begin{figure}
\centering 
\includegraphics[width=1\linewidth]{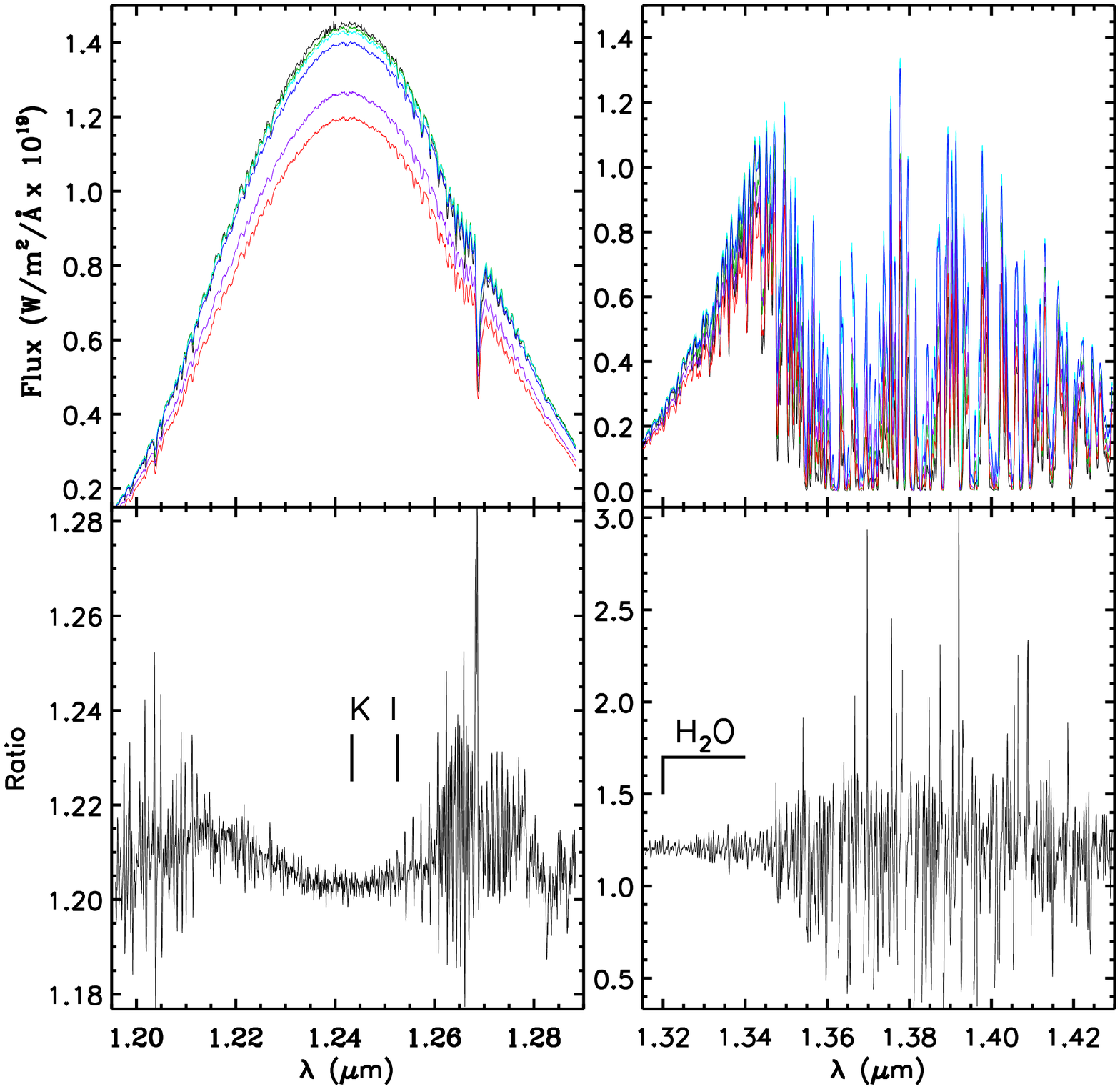}
\caption{\footnotesize \textit{Top panels:} Extractions from the six spectra of the HD~98042 telluric standard  observed on 23 February 2014 for the orders that contain the 1.24--1.25\micron \ion{K}{1} doublet and the 1.32\micron\ H$_2$O band head.  \textit{Bottom panels:} Ratios of the brightest to faintest telluric standard spectra show no above-noise systematics near the \ion{K}{1} lines and the H$_2$O band head.}
\label{fig:tells}
\end{figure}

To verify that the changes in the line strengths between minimum and maximum light are not caused by residual telluric effects, we also investigated changes in the telluric spectra.
The top panels of Figure~\ref{fig:tells} show the six spectra of the same telluric standard (HD~98042) observed on the first night (23 February 2014) for the orders that contain the 1.24--1.25\micron \ion{K}{1} doublet and the 1.32\micron\ H$_2$O band head. We see that the overall flux was indeed changing over the course of the night. Because we divide Luhman 16B by Luhman 16A in our assessment of the flux variability, this telluric variability is removed.
Furthermore, we see that the ratios of the brightest to the faintest telluric standard spectra in either order are relatively flat, with no high-frequency features above the noise level at the wavelengths of the \ion{K}{1} doublet or the H$_2$O band head
(lower panels of Figure~\ref{fig:tells}). We do notice changes in the telluric \ion{Na}{1} feature at 1.268\micron, but these do not affect other absorption features intrinsic to the brown dwarf. We conclude that although the recorded flux from the telluric standard was changing throughout the night, it did not affect our results.

\subsection{Variability and Absorption Strength Changes}\label{sec:var}

Because our spectra span a relatively wide wavelength range (0.9--2.5\micron), we are able to assess the flux variability over a broad range of atmospheric levels (e.g. \citealp{morley14,karalidi16}). The resolution of our spectra (R$\sim$4000) also allows us to accurately determine how each of the absorption features changes as the emergent flux varies.

Based on the first trough and peak of the first night of data in Figure \ref{fig:lightcurve_ratio}, it appears that the variability amplitude for the continuum next to the \ion{K}{1} absorption lines is larger in the NIR than it is in the optical. The NIR amplitude is $\sim$16\% peak-to-trough while the optical amplitude is $\sim$12\% peak-to-trough.  Uncertainties in the amplitude are on average $\pm$1\%. The continuum flux changes by similar amounts next to the other dominant absorbers---FeH (0.975--0.985$\micron$ continuum), \ion{Na}{1} (1.145--1.155\micron\ continuum) and water (1.280--1.322$\micron$) continuum---by 15\%, 15\% and 13\%, respectively.  The variation appears to decrease in the $K$-band continuum next to the CH$_4$ features. The 2.190--2.200$\micron$ continuum next to the 2.20$\micron$ CH$_4$ band varies by 6\% and the 2.300--2.310$\micron$ continuum next to the 2.31$\micron$ CH$_4$ band head has a large scatter rather than a clear periodic signature. 

From the ratio of the bright and faint states (Fig.\ \ref{fig:fracdiff_n1}), we find that the amplitude of the variability within the water band is only about 88\% of the amplitude of the continuum next to it.  This is as expected, since gas-phase water absorption occurs at altitudes above the cloud deck in early T dwarfs  \citep{apai13, yang15}. Conversely, the strengths of most other absorbers do not change by measurable amounts, except for small changes in the equivalent widths of \ion{K}{1} doublets.  The shorter-wavelength doublet changes by 0.18$\pm$0.08\AA\ and 0.15$\pm$0.07\AA, and the longer-wavelength doublet by 0.21$\pm$0.06\AA\ and 0.28$\pm$0.06\AA. These variations are 1--2\% stronger than in the surrounding continuum: for an overall variability amplitude of 17--18\% in the lines. Relative to the surrounding continuum, the \ion{K}{1} lines are weaker in the bright state than in the faint state.  We discuss the meaning of these results in the next section.

We sought to confirm the changes in the \ion{K}{1} lines by means of Principal Component Analysis (PCA), as done for time-resolved HST/WFC3 grism spectroscopy of L and T dwarfs by \cite{apai13} and \cite{buenzli15b}. PCA can isolate the main contributing factors in the wavelength-dependent light curve, and can potentially pinpoint the effect in the \ion{K}{1} lines as a separate component.  However, because the changes in the \ion{K}{1} line strengths are well below the noise level in the individual spectra, this approach did not result in useful information.  In addition, because we analyze all spectral orders after flattening the continuum---to avoid slit loss and telluric systematics---the PCA does not even reveal color changes in the continuum: seen as a wavelength dependence in the variability amplitude of the first principal component in the HST/WFC3 spectra of \cite{buenzli15b}.  Hence, we do not use PCA further.

\section{Discussion}

Many ideas have been put forth for the cause of the variability in brown dwarfs, Luhman 16B in particular. \cite{buenzli15a,buenzli15b} and \cite{karalidi16} invoke clouds to explain the change in flux in Luhman 16B, which is the favored mechanism for variability at the L/T transition \citep{ackerman01, burgasser02}. 
\cite{robinson14} suggest that brightness temperature variations can also be produced by temperature perturbations introduced at the base of the atmosphere.  Separately, \cite{tremblin16} suggest that in L/T transition objects, a reduced temperature gradient and temperature fluctuations caused by a thermochemical instability can reproduce the observed red colors and variable fluxes.  The empirical evidence favors a joint effect of changing cloudiness and temperature at the L/T transition.  Thus,  \cite{apai13} note that the NIR photometric and color variations of the two early-T dwarfs 2MASS J21392676+0220226 (2M2139) and SIMP J013656.5+093347 (SIMP0136) can be explained as linear combinations of cloud thickness variations and effective temperature.  

We note that any change in temperature or cloudiness of a single object may also be potentially reproducible as a spectral type change.  Such a supposition offers an attractive premise for explaining the complex spectroscopic signature of what are often referred to as spectral binary ultra-cool dwarfs: candidate unresolved doubles with components of different spectral types that show the combined spectroscopic signature of both components \citep{burgasser07}.  Indeed, 2M2139 is the best known example of an object that was initially suggested to be a strong spectral binary candidate \citep{burgasser10}, and was subsequently identified as a high-amplitude variable without evidence for a companion \citep{radigan12}. We also note that the \cite{apai13} conclusions about effective temperature and cloudiness changes as drivers for variability are not inconsistent with excursions along the spectral type sequence.

We explore the phenomenological characterization of variability in the context of spectral type changes in the following sections.

\subsection{Spectral Type Changes
\label{sec:spt_changes}}

If the observed variations were the result of changing spectral morphology, one would expect the changes in absorption strengths to follow a spectral sequence. The ratio of an earlier spectral type representing the bright state and a later spectral type representing the faint state might be able to explain the wavelength-dependent differences in Figure \ref{fig:fracdiff_n1}. The spectral type change would be a small one considering that the changes in absorption strengths between spectral subtypes can be as small as only a few percent.  

We investigated the amount of spectral type change necessary to reproduce the bright-to-faint ratios for Luhman 16B for some of the main absorption features in Figure \ref{fig:fracdiff_n1}. We created template spectra for the bright and faint states by combining pairs of L8 to T2 standard spectra from the BDSS survey \citep{mclean03}, with weights in increments of 0.1.  The spectrum for a bright state would be, e.g., 0.9$\times$L8 + 0.1$\times$T2, indicating proportional contributions from an L8 type and a T2 type atmosphere. We analyzed these template spectra in the same manner as we have our observed spectra, namely by forming ratios that represent bright and faint states. The spectra from the BDSS survey do not span the full NIR wavelength range---only 1.14--2.30\micron---so we were only able to assess the behavior of the $J$-band \ion{K}{1}, 1.34\micron\ H$_2$O, and 2.2\micron\ CH$_4$ features. 

\begin{figure*}
\centering
\includegraphics[scale=0.375]{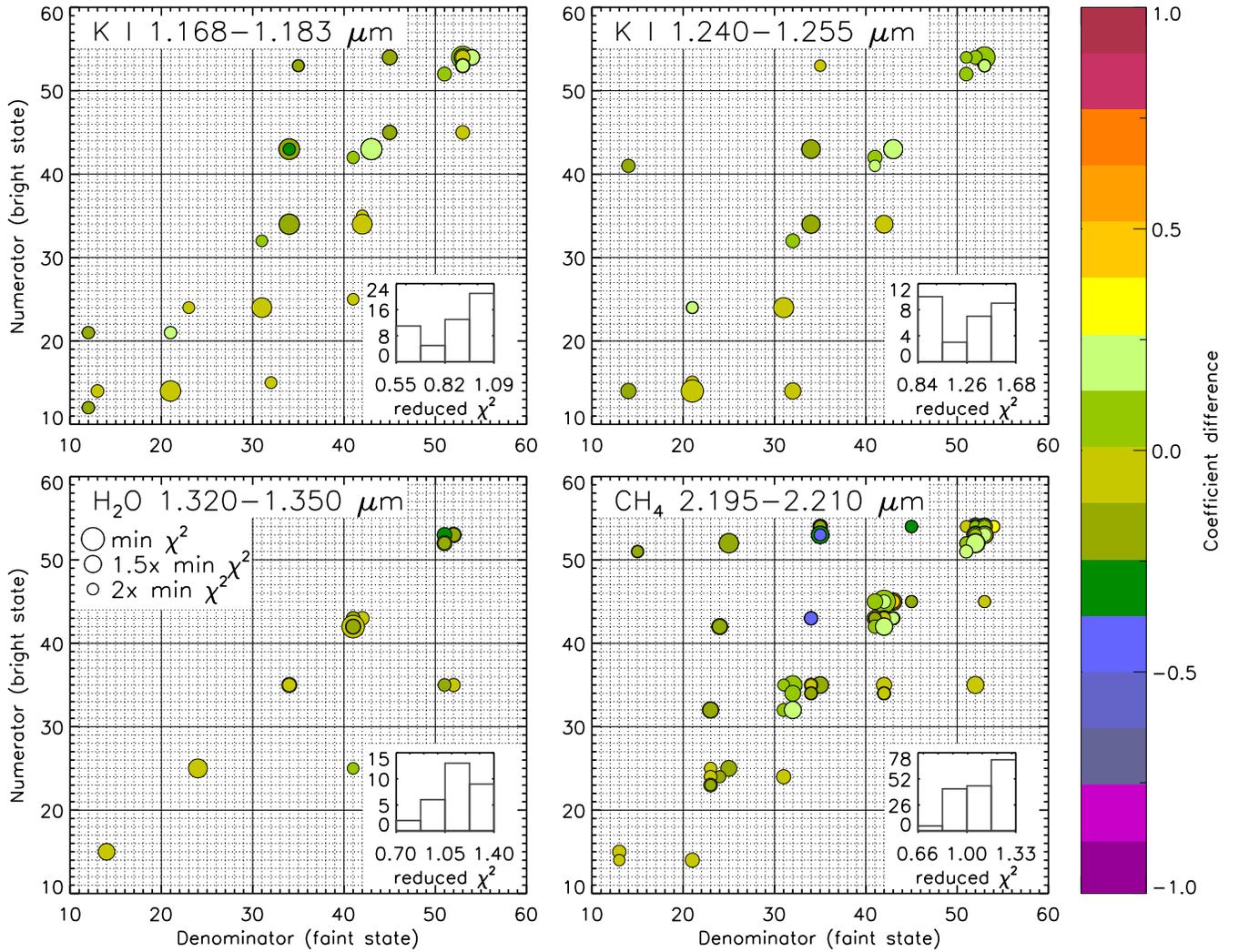}
\caption{\footnotesize Results of fitting spectral templates to our bright-to-faint state ratios of Luhman 16B for each of the four testable wavelength regions. We include results with reduced $\chi^2$ value up to only twice the minimum. Each spectral template combination is represented by two two-digit numbers: one for the numerator (indicated along the ordinate axes) and one for denominator (indicated along the abscissas).  Each digit corresponds to a spectrum that comprises the template: the first digit to the dominant spectrum in the template (carrying a $\geq$0.5 weight), and the second digit to the secondary spectrum. The greater the digit, the warmer the spectrum: L8=5, L9=4, T0=3, T1=2, T2=1. A data point on any of the panels thus reflects a combination of BDSS spectral templates that represent the numerator and the denominator in the maximum-to-minimum light ratio.
The colors correspond to the difference in the coefficient in front of the leading spectrum between the bright and faint states, with steps in units of 0.1. The sizes of the data points are inversely proportional to the reduced $\chi^2$ value, so more probable fits are shown with larger symbols. As an example, the data point at (21,14) in the top left panel has a faint state that is made of a T1 + T2 template (encoded as `21'), a bright state spectrum that is made of an T2 + L9 template (encoded as `14') and the leading coefficients differ by 0. The overall ratio of this example template is (0.9$\times$T2 + 0.1$\times$L9) / (0.9$\times$T1 + 0.1$\times$T2).  The inset histograms show the distribution of the reduced $\chi^2$ values of the best-fit templates up to twice the minimum $\chi^2$ value. }
\label{fig:chisq}
\end{figure*}

We used a reduced $\chi^2$ statistic to assess the quality of the template fits to the bright-to-faint ratios in each of the four testable wavelength regions. The errors in our reduced $\chi^2$ calculation were estimated from the scatter in the continuum regions of the bright-to-faint ratio spectra. In Figure \ref{fig:chisq} we show the best-fit combinations of spectral templates, along with the distribution of the reduced $\chi^2$ values up to twice the minimum $\chi^2$. From the top two panels of Figure~\ref{fig:chisq}, we see that the majority of the best fitting templates in the \ion{K}{1} regions are made of similar components in the bright and faint states that differ in their leading coefficients by only 0.1 to 0.2. This means that only a small change in spectral types is needed to reproduce the change in \ion{K}{1} absorption strength. The best fitting templates in the H$_2$O region are also comprised of similar spectral types in the bright and faint states.  However, while the dominant components tend to be the same, the secondary components in the best fitting templates often differ by one spectral subtype. That is, in comparison to the \ion{K}{1} regions, the change in the H$_2$O band head represents a larger change in spectral morphology. The CH$_4$ region provides fewer constraints than the other three regions. While the majority of the best fitting templates lie along the diagonal of the diagram---meaning the bright and faint states have similar dominating components---this region has a larger range of templates that fit well. This is mostly due to the fact that the bright-to-faint ratio of the CH$_4$ band head is relatively flat and is quite easy to replicate.

Overall, we see that the majority of best-fit templates cluster towards the top right of the diagram in all wavelength regions. This means that most of the best fitting templates are dominated by relatively warm L8 or L9 spectra with a smaller contribution from the cooler T1 or T2 spectra. There are a few templates that fit relatively well that are dominated by cooler T1 or T2 spectra but they do not fit well in all wavelength regions. We also see that the faint state template has a greater contribution from a cooler secondary spectrum than the bright state. 

\begin{figure*}
\centering 
\includegraphics[scale=0.35]{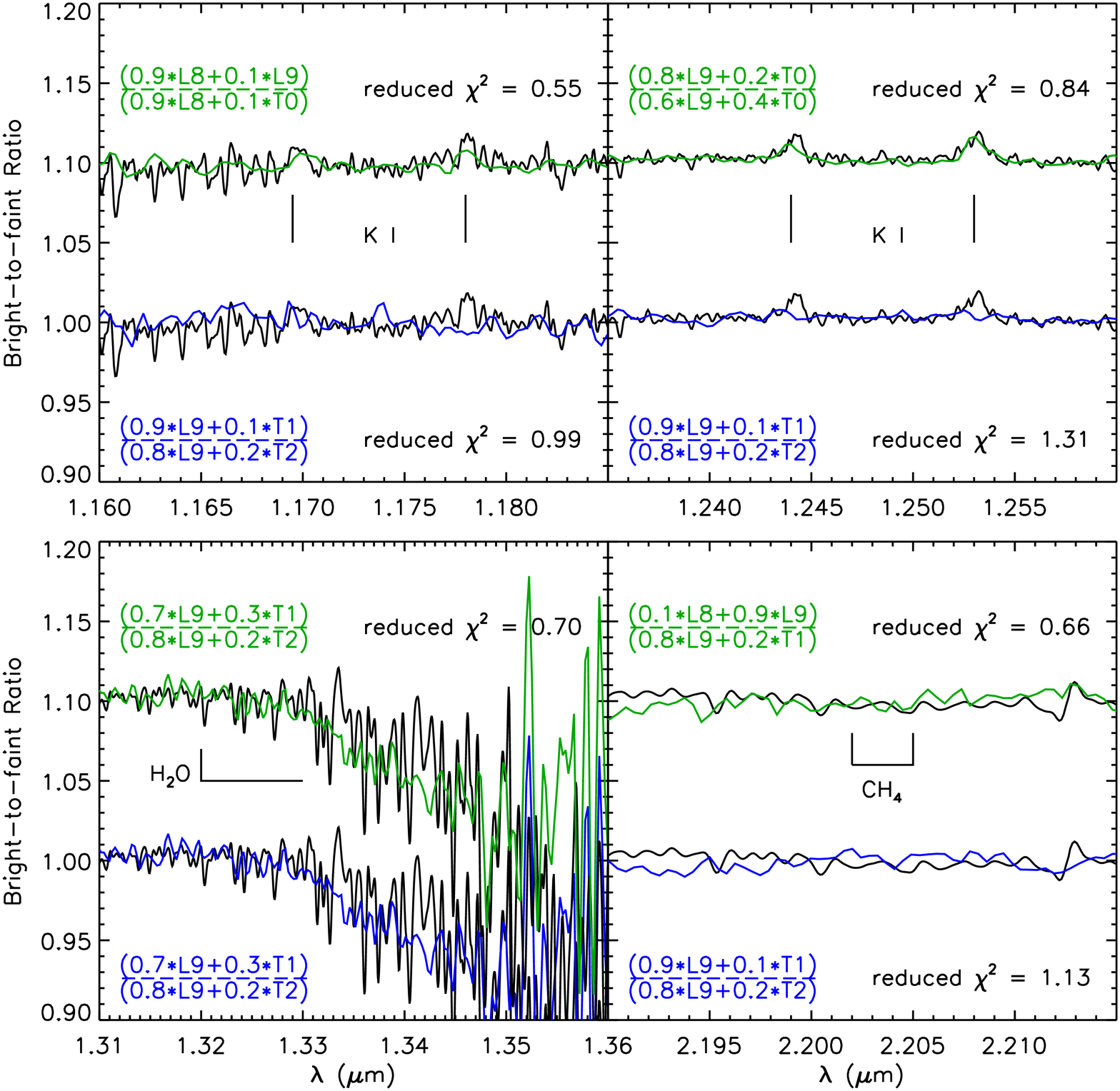}
\caption{\footnotesize Ratio of template spectra representing the bright and faint states for the two \ion{K}{1} doublets and the H$_2$O and CH$_4$ band heads created using standard spectra from the BDSS survey \citep{mclean03}. The green comparison spectra are the absolute best-fitting templates and the blue comparisons are the set of templates that have the same dominant and secondary spectra and fit well with relatively low reduced-$\chi^2$ values in all wavelength regions. In the H$_2$O region, both templates are the same. The standard spectra used to create the templates are: 2MASS J16322911+1904407 (L8), DENIS-P J025503.3--470049.0 (L9), 2MASSI J0423485--041403 (T0), SDSS J08371721--0000180 (T1) and SDSSp J125453.90--012247.4 (T2).}
\label{fig:best_rat}
\end{figure*}

The ratios of spectral template linear combinations with the minimum $\chi^2$ values for the selected absorbers are (0.9$\times$L8 + 0.1$\times$L9) / (0.9$\times$L8 + 0.1$\times$T0) and (0.9$\times$T2 +0.1$\times$L9) / (0.9$\times$T1 + 0.1$\times$T2) for the \ion{K}{1} doublets, (0.7$\times$L9 + 0.3$\times$T1) / (0.9$\times$L9 + 0.1$\times$T2) for the H$_2$O band head and (0.9$\times$L9 + 0.1$\times$L8) / (0.8$\times$L9 + 0.2$\times$T1) for the CH$_4$ band head (Figure~\ref{fig:best_rat}). We find, however, that a combination of an L9 and T1 in the bright state and an L9 and T2 in the faint state also fits all wavelength regions with low reduced $\chi^2$ values (bottom comparison spectra in Figure~\ref{fig:best_rat}). We interpret this to mean that the dominant component of the atmosphere is relatively warm and can be represented by an L9 spectrum. The changing cloud structure can then be represented by the changing secondary spectrum in the templates. During the bright state a cool T1 component contributes to the spectral morphology but when the flux decreases and goes into the faint state, the secondary component cools even further to a T2 spectral morphology. We see from Figure~\ref{fig:best_rat} however, that even though these templates have relatively low reduced-$\chi^2$ values, they do not reproduce the bright-to-faint spectra as well as the best-fitting templates. 
Because we have only used a single spectrum from each spectral type and spectral morphology ranges even within spectral types due to a variety in atmospheric structures, the templates do not fit perfectly and we can only use these templates as estimates of what components the atmosphere is made of.

From this analysis we see that the changes in spectral features of Luhman 16B can in fact be replicated by a change in spectral types. We find that the atmosphere of Luhman 16B can be represented by a predominantly warm L8 or L9 component with a smaller contribution from a cooler T1 or T2 component. Thus, we surmise that the mechanism that governs spectral appearance across the L/T transition may be the same as the mechanism that causes apparent variability in brown dwarfs. 

The preceding conclusion is borne out of the use of spectral templates as principal components.  However, the result is unusual in that while the fainter state is slightly cloudier, it is best represented by a combination of subtypes that has a greater weighting on the early-T component (Fig.~\ref{fig:best_rat}).  This is contrary to the current understanding of spectroscopic sequencing across the L/T transition, where early T dwarfs are considered to be less cloudy than late L dwarfs.  Hence, we speculate that the ordering of the L9--T1 spectral subtypes, while monotonic in CH$_4$ absorption strength, is not monotonic in effective temperature and fractional cloud coverage.   This hypothesis is independently corroborated by two other facts.  First, the L9--T1 spectral subtypes are missing from the optical late-L and T spectroscopic sequence, which jumps from L8 to T2 \citep{kirkpatrick05}, and indicates that the change in physical properties is too small to affect the red optical spectroscopic appearance.  And second, the near-IR \ion{K}{1} line strengths, which are sensitive to temperature and clouds in addition to surface gravity, follow a random progression in the L7--T2 range \citep{faherty14}.  We therefore surmise that individual subtypes in the L9--T1 range represent random realizations of cloud fraction and effective temperature---even if only over a narrow range---and that the ordering by methane absorption strength obscures these effects.

\subsection{Cloud Tomography}

The spectroscopic differences in the variability can be combined with radiative transfer and condensation models to infer the pressure levels of the cloud decks.  If a cloud deck resides at a lower pressure level than (i.e., above) the atmospheric regions contributing mostly to the source function at a certain wavelength, e.g., in the 1.25--1.32\micron\ continuum, then the clouds will attenuate the detected flux at this wavelength.  Conversely, if in a neighboring wavelength region the source function originates at lower pressure levels, as in the 1.33--1.5\micron\ water band, then the same cloud will have little influence on the detected flux.  The combined effect across the 1.25--1.5\micron\ region will be that of higher-amplitude variability in the continuum and the lower-amplitude variability in the water band: as found in \citet{apai13}, and as we confirm in Figure \ref{fig:lightcurve_ratio}.

With spectroscopic monitoring across the NIR, and at sufficiently high dispersion to resolve the pressure-sensitive alkali line profiles, we now have ample information to accurately map the vertical cloud structure in Luhman 16B.  We conduct a preliminary analysis of the wavelength-dependence of the variability here, with more detailed inverse modeling deferred to a later publication.

We first consider the continuum variations, away from dominant atomic or molecular absorbers.  From Figure \ref{fig:lightcurve_ratio}, we see that the continuum variations decrease roughly monotonically with decreasing pressure level or increasing altitude \citep[cf.\ Figure 12 in][]{karalidi16}. In order of decreasing pressure, the flux varies by 15\% (FeH continuum), 15\% (\ion{Na}{1} continuum), 16\% (NIR \ion{K}{1} continua), 13\% (H$_2$O continuum), 12\% (optical \ion{K}{1} continuum), 6\% (2.2\micron\ CH$_4$ continuum), and $\sim$2\% (2.31\micron\ CH$_4$ continuum). The decreasing variation with increasing altitude suggests that the optically thick clouds are located deep in the atmosphere and only directly affect the flux originating in the lower atmosphere. The peak in amplitude at the 1.23\micron\ continuum (near \ion{K}{1}) indicates that the clouds reside nearest this pressure level and have a decreasing affect on the flux originating above it.

We are the most interested in how the absorption features change over the course of the observations. From the bright-to-faint ratio of Luhman 16B, we see that there is no discernible change in strength of the FeH, \ion{Na}{1} or CH$_4$ absorption. The only features which have a measurable change in strength relative to the surrounding continuum are the H$_2$O and \ion{K}{1} features.  In the case of the H$_2$O absorption, we observe a similar behavior as seen by \citet{apai13} and \citet{yang15} in early T dwarfs.  Namely, the strength of the water absorption is greater relative to the continuum in the bright state compared to the faint state.  \citet{yang15} explain this by a high-altitude water haze, which resides above the inhomogeneous cloud layer driving variability in early T dwarfs.

Unexpectedly, the bright-to-faint ratios of the \ion{K}{1} absorption features are greater than unity. This means that the absorption strength relative to the continuum in the faint state is greater than it is in the bright state. If we assume the most intuitive scenario of a cloud obscuring the deeper layers in the faint state, then with part of the \ion{K}{1} column density removed, we would expect the lines to decrease in strength. Instead, we observe the opposite. That is, when the cloud comes in and obscures the flux coming from the deeper atmosphere, the NIR \ion{K}{1} absorption increases.

We consider two possible explanations for this phenomenon.  On one hand, we may be observing a layer of enhanced neutral potassium haze that exists only above the cloud deck.  Alkali cloud hazes have successfully been created in Earth's upper stratosphere by injecting sodium atoms with rockets (e.g., \citealp{blamont01}).  In the stable layers of Earth's upper stratosphere these clouds have relatively long lifetimes.  The conditions of a stratosphere above the dominant cloud deck in an ultra-cool atmosphere could be similarly stable. We note that at the lower end of the $\approx$1000--1800~K brightness temperatures probed by the 1--6$\micron$ continuum in early T dwarfs \citep{morley12}, and at the 1--5 bar pressures that correspond to the peak of the 1.0--1.3$\micron$ source function in Luhman 16B \citep[see, e.g., Figure~12 in][]{karalidi16}, neutral potassium is close to equilibrium with its chloride, KCl \citep[see Figure 14 in][]{burrows01}.  It is conceivable that neutral potassium can therefore be locally enhanced through the dissociation of KCl.  However, this runs in the opposite sense of what may be expected.  The temperature above the cloud top should be lower than if the optically thick clouds were absent, as they impede the radiative heat transfer from the deeper atmosphere.  Hence, we would expect enhanced formation of KCl above the cloud deck, rather than enhanced dissociation, and so depletion of neutral potassium through this mechanism.

On the other hand, rather than changing neutral potassium abundances, the varying strengths of the \ion{K}{1} doublets could simply reflect the different states of the temperature-pressure relation in the presence vs.\ absence of a cloud deck.  When a cloud deck is present, the observable emission from the atmosphere is cooler.  As discussed in Section~\ref{sec:spt_changes}, the observed spectroscopic variability can be well represented by increasing contributions from cooler spectral templates in the faint state. Considering the NIR spectroscopic sequences of L and T dwarfs from \cite{kirkpatrick05} and \cite{cushing05}, we see that the absorption strength of \ion{K}{1} decreases from early- to late-L dwarfs until $\sim$L8. However, between the L8 and T2 spectral types, the strength increases before it decreases again into the mid- to late-T dwarfs and disappears altogether. This agrees with the observed 1--2\% stronger \ion{K}{1} absorption in the cooler state of Luhman 16B.

Although we cannot conclusively determine the specific mechanism that causes the increased absorption strength---changing \ion{K}{1} abundances or T-P profiles---it is evident that the effects of the variability can be reproduced by changing combinations of spectral types. Thus, the mechanism that is responsible for the NIR photometric variability and the change of the \ion{K}{1} absorption strengths may well be identical to the one that governs the diversity of spectral morphology across the L/T transition.

\vskip 0.5in

\section{Conclusions}

We monitored Luhman 16AB for two nights at $R\sim4000$ over the 0.9--2.5\micron\ wavelength range.  This experiment has offered the unprecedented opportunity to study the atmospheric changes in the more highly variable early T-type secondary component at high SNR. Our analysis shows that: 
\begin{itemize}
\item the variability amplitude of Luhman 16B decreases with decreasing pressure level, indicating that the variability mechanism operates at pressures at least as high as those corresponding to the 1.2\micron\ continuum ($\sim$2-3 bars; \citealp{karalidi16});
\item the NIR \ion{K}{1} absorption strength increases in the faint state, either because of an increased \ion{K}{1} abundance above the cloud deck or because of changes in the temperature-pressure profile;
\item the variability behavior can be decomposed as a changing combination of late-L and early-T spectral templates. 
\end{itemize}

The decreasing variability with pressure confirms the results of \cite{apai13} which demonstrates that the variability causes the $J$-band flux to vary more than the $J-K_s$ color. This gives the L/T transition a wide range in $M_J$ magnitudes on a color-magnitude diagram. We have been able to reproduce the bright-to-faint ratio of Luhman 16B with a template which is predominantly a relatively warm L8 or L9 with a smaller contribution from a cooler T1 or T2 component. The faint state also has a cooler secondary component than the bright state meaning that during the faint state, part of the atmosphere is slightly cooler than in the bright state. 

We interpret the ability to reproduce the ratios with changing spectral templates as an indication that the governing mechanism of brown dwarf variability is the same as the one responsible for the diversity in spectral morphology across the L/T transition.  We also find that since the fainter, cloudier state paradoxically requires greater contributions from clearer T-type photospheres, the L9--T1 portion of the spectral sequence does not follow a monotonic trend in either effective temperature or cloud coverage.

\acknowledgements

This work was supported by an NSERC Discovery grant to S.M.\ at the University of Western Ontario. The FIRE observations reported herein were made under Chilean program CN2014A-78. This project is supported by the Ministry of Economy, Development, and Tourism�s Millennium Science Initiative through grant IC120009, awarded to The Millennium Institute of Astrophysics, MAS. This work was performed in part under contract with the California Institute of Technology (Caltech)/Jet Propulsion Laboratory (JPL) funded by NASA through the Sagan Fellowship Program executed by the NASA Exoplanet Science Institute. 

\facility{Magellan:Baade (FIRE)}
\software{IDL, FIREHOSE}

\bibliographystyle{apj}

\end{document}